# Spin Dynamics in Driven Composite Multiferroics

Zidong Wang (王子东) and Malcolm J. Grimson
*Department of Physics, the University of Auckland, Auckland 1010, New Zealand*
*E-mail address: Zidong.Wang@auckland.ac.nz*

A spin dynamics approach has been used to study the behavior of the magnetic spins and the electric pseudo-spins in a 1-D composite multiferroic chain with a linear magneto-electric coupling at the interface. The response is investigated with either external magnetic or electric fields driving the system. The spin dynamics is based on the Landau-Lifshitz-Gilbert equation. A Gaussian white noise is later added into the dynamic process to include the thermal effects. The interface requires a closer inspection of the magneto-electric effects. Thus we construct a 2-D ladder model to describe the behavior of the magnetic spins and the electric pseudo-spins with different magneto-electric couplings.

## I. INTRODUCTION

The materials exhibiting more than one ferroic (magnetic, electric or elastic) state are known as multiferroics [1]. A composite multiferroic material is defined as the heterostructure in ferroelectric (FE) and ferromagnetic (FM) orders. The FM part is a normal magnetic metal (e.g. iron, cobalt or nickel), whereas the FE part is, for instance, $BaTiO_3$ or $PbTiO_3$. Recently, this type of materials has received much theoretical [2-10] and experimental [11,12] investigation, due to induce electric polarization (magnetization) by applying a magnetic (electric) field. This phenomenon is called the magneto-electric (ME) effect, it was first discovered by P. Curie in 1894 [13]. The key to understanding the nature of the ME effect in composite multiferroic materials is the knowledge about a coupled magnetic and electric response by elastic interaction [14-16]. For a magnetized FM material, it can produce the shape deformation due to the magnetostriction. The deformation acts as a mechanical strain, it impacts the coupled FE material in form of stress, resulting in an induced polarization in FE material due to the piezoelectric effect. Vice versa, the ME effect is driven by the electric side. The ME effect only occurs at the interface between FM and FE. In this paper, our simulations assume the ME effect is controlled by a linear ME coupling as the mediator of elastic interaction. The performance of the responses of the driving and the driven parts is the purpose of this paper.

In order to study this phenomenon theoretically, we introduce a spin dynamics approach with an amount of magnetic spins and locations of electric dipoles in a 1-D FM/FE composite multiferroic chain, shown in Fig. 1. Interestingly a dynamic simulation was also introduced by D. W. Wang, *et.al* for the study of single phase multiferroic materials (i.e., $BiFeO_3$) [17]. The novelty here is that we use a pseudo-spin model to represent the electric dipoles in the spin dynamics [18]. The pseudo-spin model was first conjectured by P. G. de Gennes in 1963 [19], for calculating the energy of the proton system stored in the order-disorder ferroelectric crystals, while the term 'pseudo-spin' was coined by R. J. Elliott, *et.al*, in 1970 [20]. Later, W. Zhong, *et.al.*, developed a first-principles approach to study the structural phase transitions and finite temperature properties in perovskite FE materials [21,22]. This approach can be applied to $BaTiO_3$, $PbTiO_3$, $KNbO_3$, etc. The effective Hamiltonian was used to carry out Monte Carlo simulations and determine the phase transformation behavior.

In this paper, the effective Hamiltonians for FM and FE have been introduced in section II. The technical details of the spin dynamics simulation are presented in section III. In section IV, the results of the magnetic driven and the electric driven systems are obtained. In section V, an influence of thermal agitation has been considered in spin dynamics process. The thermal fluctuations are represented by adding a stochastic field, as mentioned by W. F. Brown in 1963 [23]. In section VI, we construct a 2-D ladder model, in order to take a closer inspection of the motions of the magnetic spin and the electric pseudo-spin with different ME couplings. Conclusions are drawn in section VII.

## II. MODEL

For this purpose, we consider a 1-D multiferroic composite chain, which coupled by FM and FE parts. Each part consists of fixed number of magnetic spins and electric pseudo-spins. The FM/FE chain is glued by the ME coupling at the interface of the last magnetic spin and the first electric pseudo-spin. The schematic view is in Fig. 1.

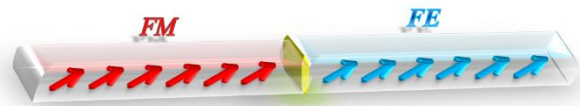

FIG. 1. Schematic of the 1-D composite multiferroic (FM/FE) chain, with the red arrows are magnetic spins and the electric pseudo-spins are blue arrows. The interface with the ME coupling is given in a yellow wall.

The total energy of the composite multiferroic system in general consists of three parts,

$$\mathcal{H} = \mathcal{H}_{FM} + \mathcal{H}_{FE} + \mathcal{H}_{ME} \quad (1)$$

$\mathcal{H}_{FM}$ is the conventional Heisenberg Hamiltonian [24] describes the FM part with $N$ magnetic spins,

$$\mathcal{H}_{FM} = -J_{FM}\sum_{\langle i,j \rangle}^{N}\left(\vec{S}_i \cdot \vec{S}_j\right) - K\sum_{i=1}^{N}\left(S_i^z\right)^2 - H(t)\sum_{i=1}^{N}S_i^z \quad (2)$$

where $J_{FM}$ is the nearest-neighbor exchange interaction coupling and K is the z-directional uniaxial anisotropy coefficient. The magnetic spin vector, $\vec{S}_i = \left(S_i^x, S_i^y, S_i^z\right)$, at site $i = 1,...,N$, with the normalization $\|\vec{S}_i\| = 1$, and the notation $\langle i, j \rangle$ characterizes that the sum is restricted to nearest-neighbor pairs of spins, each pair being counted only once. The last term in Eq. (2) shows the Zeeman energy induced by the magnetic spins and an external magnetic field $H(t)$ in sinusoidal type [25],

$$H(t) = H_0^z \sin(\omega t) \quad (3)$$

this field has been applied in the z-direction only with the time dependent form.

The energy in the FE part, $\mathcal{H}_{FE}$, is described by an Ising model in a constant transverse field. The first introduction of transverse Ising model (TIM) was by de Gennes [19] in 1963, to study the phase transition in order-disorder and KDP-type ferroelectrics. Thus, $\mathcal{H}_{FE}$ is the Hamiltonian of TIM with $N$ electric pseudo-spins that represents the interacting dipoles [18,26],

$$\mathcal{H}_{FE} = -\Omega^x \sum_{i=1}^{N} P_i^x - J_{FE}\sum_{\langle i,j \rangle}^{N}\left(P_i^z P_j^z\right) - E(t)\sum_{i=1}^{N} P_i^z \quad (4)$$

where $P_i^x$ and $P_i^z$ are the x- and z-components of the unit vector, $\mathbf{P}_i$, represent the electric dipole moments (i.e., polarization) at site $i$. In Eq. (4), the first term is the transverse energy with a transverse field, $\Omega^x$ is applied in x-direction perpendicular to the Ising z-direction [27]. The coupling, $J_{FE}$ indicates the nearest-neighbor exchange interaction provides the interaction energy in the second term. The last term is, if the system is subject to a dynamic electric driving field in the z-direction, $E(t)$ [Eq. (5)], that couples to the pseudo-spins in the system.

$$E(t) = E_0^z \sin(\omega t) \quad (5)$$

The interfacial energy, $\mathcal{H}_{ME}^{chain}$ in the FM/FE chain, between the last magnetic spin and the first electric pseudo-spin are described by the dipole-spin interaction Hamiltonian in Eq. (6), with a linear ME coupling, $g$ [2-5].

$$\mathcal{H}_{ME}^{chain} = -g\left(\mathbf{S}_N \cdot \mathbf{P}_1\right) \quad (6)$$

## III. SPIN DYNAMICS

To describe the dynamics of magnetic spins and electric pseudo-spins, an equation named Landau-Lifshitz-Gilbert (LLG) equation in the absence of the thermal agitation, has been used at the atomic level [4,28-30]. Eq. (7) shows the time evolution of the spin response (i.e. magnetization) in the FM part,

$$\frac{\partial \mathbf{S}_i}{\partial t} = -\gamma_{FM}\left[\mathbf{S}_i \times \mathbf{H}_{S_i}^{eff}(t)\right] - \lambda_{FM}\left[\mathbf{S}_i \times \left[\mathbf{S}_i \times \mathbf{H}_{S_i}^{eff}\right]\right] \quad (7)$$

Where $\gamma_{FM} = \frac{\gamma_0}{1+\alpha_{FM}^2}$, $\gamma_0$ is the gyromagnetic ratio relating the magnetization to its angular momentum, and $\alpha_{FM}$ is the dimensionless damping factor. $\lambda_{FM} = \frac{\gamma_0 \alpha_{FM}}{1+\alpha_{FM}^2}$ denotes the Gilbert damping term. The magnetic effective field, $\mathbf{H}_{S_i}^{eff} = -\delta\mathcal{H}/\delta\mathbf{S}_i$ is the derivative of the system Hamiltonian of Eq. (2) with respect to the magnetization, acting on each magnetic spin.

In the FE part, we use a simple pseudo-spin model to describe the locations of the electric dipole. Since an electric dipole is a separation of positive and negative charges, a measure of this separation gives the magnitude of the electric dipole moment, it is a scalar. In the spin dynamics approach, no precession of the pseudo-spins is expected (i.e. $\gamma_{FE}' = 0$) and the dynamic responses of electric pseudo-spins (i.e. polarizations) are described [18,31,32],

$$\frac{\partial \mathbf{P}_i}{\partial t} = -\lambda_{FE}\left[\mathbf{P}_i \times \left[\mathbf{P}_i \times \mathbf{H}_{P_i}^{eff}\right]\right] \quad (8)$$

where $\lambda_{FE}$ is the intrinsic damping parameter for the electric pseudo-spins, and the electric effective fields for the pseudo-spins, are defined as a functional derivative of Eq. (4), $\mathbf{H}_{P_i}^{eff} = -\delta\mathcal{H}/\delta\mathbf{P}_i$. The electric effective fields, without any y-component, only have a constant transverse field in the x-component. Thus, the polarization in the z-direction dominates the motion of electric pseudo-spins.

Introducing the dimensionless variables with $\mathbf{s}_i = \mathbf{S}_i / S_0$ and $\mathbf{p}_i = \mathbf{P}_i / P_s$, given the normalized LLG equations,

$$\frac{\partial \mathbf{s}_i}{\partial t} = -\gamma_{FM}^*\left[\mathbf{s}_i \times \mathbf{h}_{S_i}^{eff}(t)\right] - \lambda_{FM}^*\left[\mathbf{s}_i \times \left[\mathbf{s}_i \times \mathbf{h}_{S_i}^{eff}\right]\right] \quad (9)$$

and

$$\frac{\partial \mathbf{p}_i}{\partial t} = -\lambda_{FE}^*\left[\mathbf{p}_i \times \left[\mathbf{p}_i \times \mathbf{h}_{P_i}^{eff}\right]\right] \quad (10)$$

where $\mathbf{h}_{S_i}^{eff} = -\frac{1}{\mu}\frac{\delta\mathcal{H}}{\delta\mathbf{s}_i}$ and $\mathbf{h}_{P_i}^{eff} = -\frac{1}{\epsilon}\frac{\delta\mathcal{H}}{\delta\mathbf{p}_i}$ are the reduced effective fields.

## IV. RESULTS

To demonstrate the responses of magnetic spins and electric pseudo-spins to a magnetic/electric driving field using the spin dynamics approach. The number of magnetic spins and electric pseudo-spins are set at $N = 50$ each, in order to clearly observe the behaviour in driven part. The other parameters with a '*' label are dimensionless. Large couplings of the nearest-neighbour exchange interaction $J_{FM}^* = J_{FE}^* = 50$, and the ME coupling $g^* = 50$ has been used to enhance the performance of the energy transition. The coefficient of anisotropy is $K_{FM}^* = 0.1$ in the FM part. The transverse field in the FE part is $\Omega_{FE}^* = 0.1$. In LLG equations, $\gamma_{FM}^* = 1$ and the same damping coefficients $\lambda_{FM}^* = \lambda_{FE}^* = 0.1$. The dimensionless external driving field is either magnetic [Eq. (3)] or electric [Eq. (5)] in each run, with fixed amplitude of $H_0^* = 10$ or $E_0^* = 10$ in the $z$-direction only. The switching frequency is set by the switching field $\omega^* = 0.01$. This means $\lambda^*/\omega^* = 10$, the stiffness characterises an appropriate response of the system. Free boundary conditions have been applied, and the initial states have been set at random. The numerical results show that the $z$-component of magnitudes of the magnetisation $S_z$ and the polarisation $P_z$, obtained by a fourth order Runge-Kutta method with a time step $\Delta t = 0.0001$.

We apply a dynamic electric field to the FM/FE chain. Generally, the mean electric polarization can form a hysteresis loop with the driving field seen in Fig. 2(b). Since, the FM part couples with the FE, the responses of the magnetic spins are driven by the variation of the electric polarization. Thus a relatively weak hysteresis loop of the mean magnetization and the driving field can be obtained in Fig. 2(a). For a closer inspection of the instantaneous responses of each individual spin/pseudo-spin to the dynamic driving field, we select seven specific cases expressing in different symbols are depicted in Fig. 2(c). The left hand side shows the magnitudes of the magnetic spins in the $z$-component [red], and the right hand side shows the magnitudes of the electric pseudo-spins [blue], the interface between FM and FE is represented by a yellow line. The process starts from the '△' chain at the bottom. The electric pseudo-spins give quick response due to they are driven directly by the driving field is indicated in the second '∗' and third '□' chains. Then the next three chains show the magnetic part catch up slowly. The average values of each chain are indicated in Figs. 2(a) and (b) with corresponding symbols. The overview of the dynamic responses is given in Fig. 3(a) for three periods of the driving field. The multiple colors in this figure characterize the $z$-component magnitudes of the magnetization and the electric polarization at each spin-site. The delay behavior in the FM part can be observed as shown above in Fig. 2(c).

Similarly, the FM/FE chain is driven by a magnetic field and its analogous behavior is obtained in Fig. 3(b). For the driven part, the polarization is much weaker than the magnetization in Fig. 3(a). This is due to the TIM being used in electric pseudo-spin model, which only contains the $z$-component interaction energy [Eq. (4)]. It results in low energy transitivity and induces distinct decay behavior in further electric pseudo-spins. In contrast, the magnetic spins used in the classical Heisenberg model are the sum of the $x$-, $y$- and $z$-component interaction energies [Eq. (2)]. For the same parameters used in the simulation, the total energy in the magnetic spins is larger than the electric pseudo-spins. Thus there is a smaller hysteresis loop for the polarization with the driving field as obtained in Fig. 4(b). A closer inspection of the dynamic response of the FM/FE chain is shown in Fig. 4(c) with six specific cases. Some fluctuations occur during the process in FM part are indicated in the '∗', '□' and '☆' chains. This causes the energy reflection from the interface, as a result of the limited capacity of energy absorption in the electric pseudo-spins with TIM. This behavior can also be observed in Fig. 4(a) as weak fluctuations in the magnetic hysteresis loop.

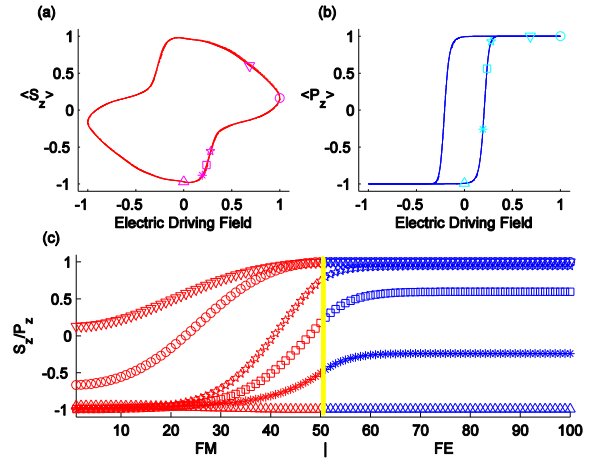

FIG. 2. (a) and (b) show the $z$-component mean values hysteresis loops of magnetization and electric polarization, respectively. (c) illustrates the responses of magnetic spins [red] and electric pseudo-spins [blue] as spin waves in seven specific cases to an **electric** driving field.

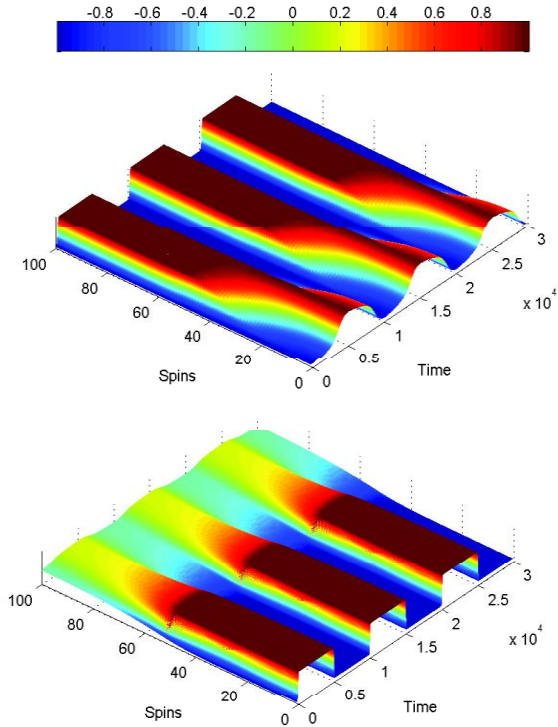

FIG. 3. This figure illustrates the *z*-component magnitudes of magnetization and electric polarization in three periods. (a) is driving by an electric field, and (b) is driving by a magnetic field.

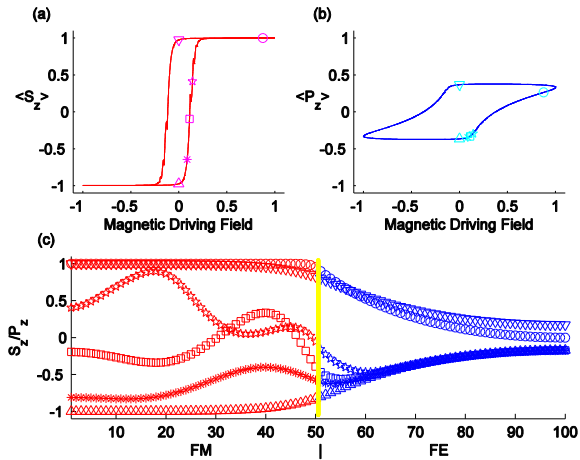

FIG. 4. (a) and (b) show the *z*-component mean values hysteresis loops of magnetization and electric polarization, respectively. (c) illustrates the responses of magnetic spins [red] and electric pseudo-spins [blue] as spin waves in six specific cases to an **magnetic** driving field.

## V. STOCHASTIC EFFECTS

The thermal agitation is neglected in section IV. In experimental research, the influence of temperature is significant. For each individual spin/pseudo-spin, the orientation of the moment is continuously changed by the thermal agitation. This problem can be approached as a simplified Brownian motion theoretically. In effect, we reduce the random forces to a purely random process as a stochastic field [23].

Based on the method in section IV, we process the simulations based on the effective fields with an additional Gaussian white noise for the stochastic field [33]. Thus the stochastic effective field,

$$H_i^{eff+stoch} = H_i^{eff} + H_i^{stoch} \quad (11)$$

where $H_i^{stoch}$ is the stochastic field at each spin-site,

$$H_i^{stoch} = \frac{1}{\sigma\sqrt{2\pi}} e^{-\frac{(x-\mu)^2}{2\sigma^2}} \quad (12)$$

where $\mu$ is the mean of distribution, $\sigma$ is the standard deviation of distribution, and $x$ is the real random variable vectors in 3 degrees of freedom. Thus, Eqs. (7) and (8) become stochastic LLG equations. We use the parameters given in section III, with a 'balanced' mean Gaussian distribution (i.e. $\mu=0$). The temperature influences are proportional to the magnitudes of Gaussian standard deviation, $\sigma$.

The numerical results are shown in Fig. 5, the top two panels are driving by magnetic field and the bottom panels are driving by electric field. We compare the noise level $\sigma = 0.3$ [Figs. 5(a) and (c)] and $\sigma = 1$ [Figs. 5(b) and (d)], the effects of thermal agitation can be clearly observed by a comparison of the noiseless cases in Figs. 3(a) and (b). For a small level of noise, $\sigma = 0.3$, the behavior of responses are almost similar to the absence of thermal agitation. Increasing the strength of noise to a moderate level, i.e., $\sigma = 1$, the thermal fluctuation plays an important role in the driven part. The Monte Carlo approach can inspect the thermal influence by its transition probability [3], but the results are qualitatively the same.

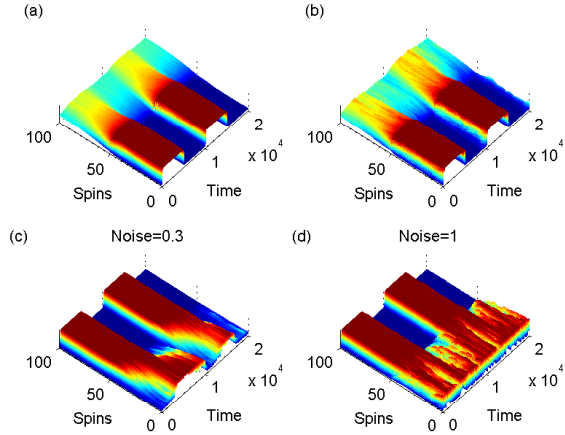

FIG. 4. Influences of thermal fluctuation in the FM/FE chain is illustrated by similar simulations as in Figs. 2 and 4, and including a Gaussian white noise with different noise levels. (a) and (b) are driven by a magnetic field; (c) and (d) are driven by an electric field.

## VI. MAGNETO-ELECTRIC EFFECTS

We have clearly identified the validity of the magnetic spin model and the electric pseudo-spin model in spin dynamic approach for composite multiferroic simulation. The preceding discussion suggests that ME effect is directly proportional to the ME coupling, $g$. It is an important factor during the energy transition from the driving part to the driven part. To demonstrate this behavior, we consider a simple 2-D ladder model to enhance the ME coupling phenomena at the FM/FE interface [9]. The ladder model with periodic boundary conditions is illustrated in Fig. 6 with $N$'s ME couplings. We use the same simulation approach as in section IV, but the interfacial energy becomes,

$$\mathcal{H}_{ME}^{ladder} = -g \sum_{i=1}^{N} (\mathbf{S}_i \cdot \mathbf{P}_i) \quad (13)$$

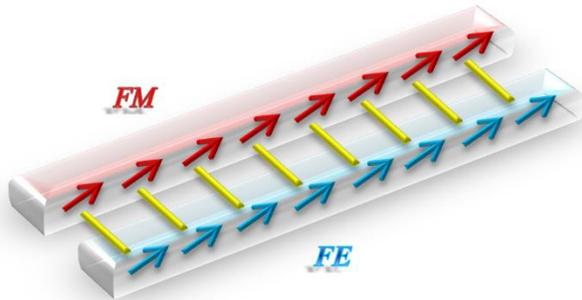

FIG. 6. Schematic of the 2-D composite multiferroic (FM/FE) ladder. The features are same as Fig. 1.

For the numerical simulations of the ladder model, we found that the maximal size tractable in reasonable computing time is given by $N=100$, and reduce the normalized nearest-neighbor exchange interactions to $J_{FM}^{*} = J_{FE}^{*} = 1$. A variation of the linear ME coupling, $g$ allows a comparison of the different responses of the magnetic spins and the electric pseudo-spins. In this section, we depict the mean magnetization/polarization hysteresis loops and a magnetic spin/an electric pseudo-spin's motions respect to the ME couplings. Results for the ME couplings of $g=1, 0.5$, and $0.2$, are shown in Fig. 7 for electric field driving and Fig. 8 for magnetic field driving.

A closer inspection for each panel in Fig. 7, the top row [Figs. 7(a)-(c)] shows the behavior respect to $g=1$, a relatively strong ME coupling. The magnetic spin trends to the direction of the transverse field, giving a similar behavior as the performance of the electric pseudo-spin in Fig. 7(b). On the other side, the electric pseudo-spin shows a homogeneous dipole motion in Fig. 7(c). Reducing the ME coupling strength, the magnetic spin gains more freedom of precession in each direction shown in Figs. 7(e) and (h), but it losses stiffness for weaker ME coupling [Figs. 7(a), (d) and (e)].

In Fig. 8, similar behaviors have been obtained as in Fig. 7, but the driving part is the magnetic spin. For a strong ME coupling, $g=1$, the precession of the magnetic spin has been limited by the electric pseudo-spin shown in Fig. 8(b). The oscillations observed in its hysteresis loop [red loop in Fig. 8(a)], indicate the axis of precession is not along to the $z$-axis. This is due to the coupling of the electric pseudo-spin to the transverse field along to the $x$-axis. The FM part gains the freedom of precession slowly with a smaller ME coupling [Figs. 8(e) and (h)], and less oscillations are shown in their hysteresis loops [red loops in Figs. 8(d) and (g)]. The electric pseudo-spin only changes its magnitude in the $z$-component [Figs. 8(c), (f) and (i)]. It has much weaker response [blue loops in Figs. 8(a), (d) and (g)] than that of the magnetic spins shown in section IV.

In a nutshell, the ME coupling in spin dynamics approach is shown to be an important role in controlling the freedom of precession of the magnetic spin. As an aside, if $g=0$, represent that driving part and driven part are absolutely isolated, no energy can be propagated. Thus, the driving part shows a general "ferro-like" behavior. But, the driven part shows a zero effect only if the ME coupling is zero.

Journal of Applied Physics **118**, 124109 (2015); DOI: 10.1063/1.4931895

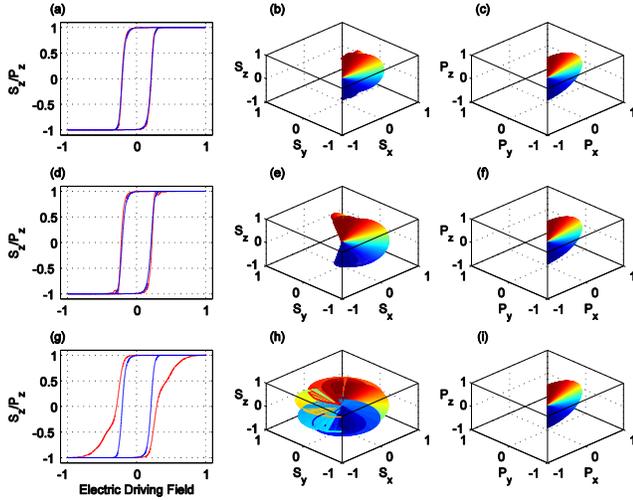

FIG. 7. This figure illustrates the trajectories of a magnetic spin [(b), (e) and (h)] and an electric pseudo-spin [(c), (f) and (i)] with different ME couplings to an **electric** driving field in the ladder model. The panels of top row with $g=1$, mid row with $g=0.5$, and bottom row with $g=0.2$. Panels (a), (d) and (g) show the z-componental hysteresis loops of the magnetic spin in red and the electric pseudo-spin in blue.

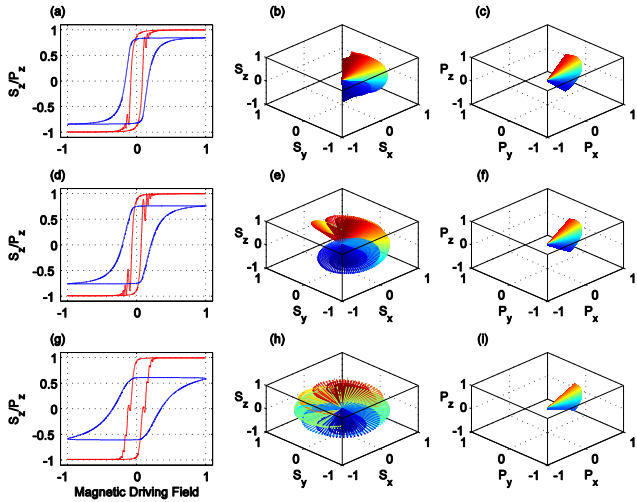

FIG. 8. This figure illustrates the trajectories of a magnetic spin (second column) and an electric pseudo-spin (third column) with different ME couplings to a **magnetic** driving field in the ladder model. The panels of top row with $g=1$, mid row with $g=0.5$, and bottom row with $g=0.2$. Panels (a), (d) and (g) show the z-componental hysteresis loops of the magnetic spin in red and the electric pseudo-spin in blue.

## VII. CONCLUSIONS

The main point of this paper is to observe the ME effects by spin dynamics on the both sides of FM and FE. A spin model has been used for magnetic spins, and an electric pseudo-spin model has been used to represent the locations of electric dipole in a precession free LLG equation. The numerical results demonstrate the different responses of the magnetic spins in the FM and the electric pseudo-spins in the FE driven by either a magnetic field or an electric field. In spin dynamics, the thermal agitation can be considered by adding a Gaussian white noise, and we use the stochastic standard deviation to control the influence of temperature. The results are qualitatively consistent with the Monte Carlo approach.

### ACKNOWLEDGEMENTS

The author gratefully acknowledges Zhao BingJin and Wang YuHua for financial support.